\documentstyle[12pt, graphics]{article}
\title{Relation between Tunneling and Particle Production in Vacuum Decay}
\author{Laura Mersini\\
\it{ Department of Physics}\\
\it{University of Wisconsin-Milwaukee}\\
\it{Milwaukee, WI 53201}\\
lmersini@uwm.edu}

\begin{document}
\date{}
\maketitle
\renewcommand{\thesection}{\arabic{section}.}
\renewcommand{\theequation}{\thesection\arabic{equation}}
\begin{abstract}
The field-theoretical description of quantum fluctuations on the background 
of a tunneling field $\sigma$ is revisited in the case of a functional Schrodinger
approach.   We apply this method in the case when quantum fluctuations are coupled to the $\sigma$ field through a mass-squared term, which is 'time-dependent' since we include the dynamics of $\sigma$ .  The resulting mode functions of the fluctuation field, which determine the quantum state after tunneling, display a ,previously unseen,  resonance effect when their mode number is comparable to the curvature scale of the bubble. A detailed analysis of the relation between the excitations of the field about the true vacuum (interpreted as particle creation) and the phase shift coming from tunneling is presented. 
\end{abstract}
\pagebreak
\section{\underline{\bf Introduction.}}
False vacuum decay through barrier penetration has been studied extensively
by a number of authors [1-10].  However there is a lot to be understood
yet about tunneling phenomena, especially if one considers its importance
as related to the very early universe.  Tunneling is often related to the
existence of a potential barrier ,but while in quantum mechanics (QM) that
barrier is in real space, in quantum field theory (QFT) one has penetration from one
field configuration to another. In QFT the effect of a tunneling potential can even be  produced
through the spatial gradient of the field. Thus the potential barriers in
$(QM)$ and $(QFT)$ have a different nature.

It is well known that tunneling in QFT is a large fluctuation or instability of
vacuum, a classically forbidden process where energies are not eigenstates
of the hamiltonian due to barrier penetration. Since particle creation is another phenomena related to vacuum fluctuation [11-18] it is of interest to probe tunneling via studying the  relation between particle creation and tunneling.

The tunneling amplitude depends strongly on $\hbar$ since it is the exponential of the negative Euclidean action divided by $\hbar$, while particle creation is insensitive to it. For more details see [7,9]. Thus, the question we ask here
is:  What is the relation between tunneling and particle creation in the
dynamic background of the bubble that forms in false vacuum decay? We will consider the functional Schrodinger equation for a quantum field $\Phi$
in a tilted quartic potential $V(\Phi)$ (fig. 1).  The field is split into a 
classical part, $\sigma$ (hereafter the tunneling field) undergoing phase
transition from false to true vacuum, and the fluctuation field, $\varphi$
around $\sigma$, giving rise to particle creation through coupling to 
$\sigma$
with a mass-squared term, $m^2(\sigma)$.The classical field $\sigma$ is defined as usual to be the one that extremizes the action. In this case, assuming an WKB ansatz for the wavefunctional, (Section 2 reviews this formalism),  is constistent with the definition of $\sigma$, since the expectation value for the lowest order WKB type solution recovers the classical solution $\sigma$ in the appropriate limit. However generally this statement would not be true particularly in the cases when there are more than one degree of freedom
  and the WKB branches are not decohered. The multidimensional tunneling in WKB formalism was initially
developed by Banks, Bender and Wu [5] and later revisited and improved
by T. Vachaspati and A. Vilenkin [6,7] as well as the Kyoto group [Hamazaki et al 8,9].

The reason for using this method is twofold:  \\
- It allows us to relate the dominant escape path,DEP [22], of $\sigma$ to the
the usual formalism of particle creation [11,12], i.e. relate the fluctuations
$\varphi$ around the DEP of the $\sigma$-field to the particles produced as
the result of tunneling.\\
- While the approach is similar to $QM$ (i.e. solving a Schrodinger equation)
there is a clear relation with the second quantization picture (shown rigorously 
by [2,3,7,8]).

The method follows closely [7,9] and we review the formalism in Sect. 2.
The ansatz of the wavefunctional is assumed to be a superposition of Gaussian
packets centered around the classical vacuum outside the barrier and around the DEP under the barrier.  One then solves the functional Schrodinger equation
for the $\sigma$-field and the fluctuation $\varphi$-field on the background
of $\sigma$, with the appropriate boundary conditions at the inner turning point.  The wavefunctional
 that describes the quantum state of the two fields,i.e.  that gives the amplitude for the fields to be in configuration {$\sigma$, $\varphi$}, 
is a 1-parameter family of solutions (the parameter can be thought of as 
a time-coordinate) and is analytically continued from under the barrier through the turning point,
to the region of the true vacuum.  The solution for the fluctuation field is given in 
Section 3.  When the solution is analytically continued, one ends up with
a squeezed state of mixed positive and negative frequencies. The coefficient of the negative frequency component than allows one to determine the particle creation number in each mode.

We find an interesting new feature for the fluctuations, namely resonance, by taking into account the dynamic of the 
background field $\sigma$. A discussion
of the results and open issues to be tackled in the future is given in
Section 4.  

All along we have assumed the $O(4)$-symmetry and the thin wall approximation.
In the appropriate limit our result recovers those obtained by the Kyoto
group [8,9]. 

\section{\underline{ Review of the Functional Schrodinger Formalism}}
The application of this formalism in connection with vacuum tunneling in QFT was initially developed by Banks, Bender and Wu [5] and
later on revisited and improved by [2,3,7,8,9].
We consider a scalar field in Minkowski spacetime with the action
\[ S = \int d^4 x \left[ \frac{1}{2} (\partial_\mu \Phi)^2 - V(\Phi) \right] \]
where
\[ V(\Phi) = \frac{\lambda}{2} (\Phi^2 - a^2)^2 - \frac{\epsilon}{2a}(\Phi + a),\quad
\frac{\epsilon}{(a^4\lambda)}\ll 1 \quad \; \]

[see Fig.1]  
\vspace*{1in}\\


Fig.1  The Potential for False Vacuum Decay via Tunneling

We want to study the fluctuation field around the dominant escape path (i.e. around
the
tunneling field $\sigma$).  Thus we split the field into the classical/tunneling field
``$\sigma$'' and fluctuations around $\sigma$, denoted by $\varphi$, i.e.
$\Phi = \sigma + \varphi$ .

Under the barrier, $\sigma$ interpolates between the 2 vacuum states by
taking the dominant escape path.  The solution for the $\sigma$-field in 
the case of potential $V(\sigma)$ is studied by 
many authors through the use of instantons in the dilute gas approximation under the barrier,
or by means of WKB and loop expansion methods outside the barrier.It has been shown that both perturbation methods (dilute gas of instantons inside and 1-loop approximation outside the barrier) match consistently at the turning point,only when the continuous symmetries of the problem
(in our case $O(4)$) are the same under the barrier and outside, where
one is finding solutions around $\sigma \simeq \sigma_\pm$. outside) [2,3].
This matching procedure can be quite complicated if one breaks different subgroups of
 the
continuous symmetries inside and outside the barrier.
\begin{itemize}
\item[{\bf 2.1 }] \underline{\bf Particle Creation as excitations of the tunneling field.}

Consider a scalar field theory $\Phi$ in a quartic potential
\[ V(\Phi) = \frac{\lambda}{2}(\Phi^2 - a^2)^2 - \frac{\epsilon}{2a}(\Phi + a),
\quad \epsilon > 0 \quad \mbox{and}\; \frac{\epsilon}{(\lambda a^4)} \ll 1 \]

As noted earlier, we split the fields in 2 parts: $\sigma$, the classical field 
$(\; \mbox{where}\; \frac{\delta S}{\delta \Phi}|_{\Phi = \sigma} = 0)$
and $\varphi$,the fluctuation field.  The false vacuum is located
at $\sigma_{_-} \simeq - \; a - \frac{\epsilon}{8 \lambda a^3} + O(\epsilon^2)$,
where $V(\sigma_{_-}) \simeq 0 + O(\epsilon^2)$.  The true vacuum is
located at $\sigma_+ \simeq a - \frac{\epsilon}{8\lambda a^3} + O(\epsilon^2)$
,where $V(\sigma_+) \simeq -\epsilon + O(\epsilon^2)$ . Then the Lagrangian
for the $(\sigma + \varphi)$-fields becomes
\begin{equation}
\;\;\;\; L = \left\{ \frac{1}{2} (\partial_\mu \sigma)^2 - V(\sigma)\right\} +
\frac{1}{2} \left\{(\partial_\mu \varphi)^2 - m^2(\sigma)\varphi^2 \right\} 
= L_\sigma + L_\varphi
\end{equation}   
where
$m^2(\sigma) = \frac{\partial^2 V}{\partial\sigma^2}$ and
$(\Box \varphi - m^2(\sigma)\varphi) = 0$
We have ignored higher order terms in the potential that would have
contributions of order $\varphi^3$ and higher. This is a consistency requirement for matching the two different perturbation methods at the turning point. When we solve the functional Schrodinger equation, the solution for the
$\sigma$-field under the barrier will contain terms up to $(\hbar^{-2})$
coming from the zeros in the determinant factor 
$K = \left(\frac{S_0}{2\pi h}\right)^2 \det^{-1/2}|- \partial_\mu \partial_\mu + V''(\sigma)|$, where ${S_0}$ is the action of the bounce in 4-dim.
These zeros are related to the 4 translational symmetries
of the problem, and it has  
been shown (see Coleman [19]) in the dilute gas
approximation that each symmetry brings a $(\hbar^{-1/2})$ factor, so
counting zeros in $K$ is equivalent to counting powers in the coupling
constants.  Further, Gervais, Vegas, Sakita, [2,3] have shown that  
for the problem at hand, expansion around the vacuum states, outside the
barrier, which in our case corresponds to $\sigma_+$, matched (analytically continued) with the solution under the
barrier is in complete agreement with the multi-instanton (dilute gas approx.)
approach [1].  Thus we can keep terms up to $\varphi^2$
in the Lagrangian and ignore higher order terms.  

We work with the functional Schrodinger equation and use the results of  Vachaspati and Vilenkin [7] for the background field $\sigma$.   
\item[{\bf 2.2}] \underline{ \bf Functional Schrodinger Equation and Solution for the Classical Field.}

From the Lagrangian in (1) we can write
\[ H \equiv H_\sigma + H_\varphi = [ \frac{1}{2} \pi^2_\sigma + V(\sigma)] +
\frac{1}{2} \pi^2_\varphi + \frac{1}{2} m^2(\sigma)\varphi^2 . \] 
The quantum state of the 2 fields is described by a wavefunctional
 $\psi[\sigma(\vec x,t), \varphi(\vec x,t)]$
which gives the amplitude for the fields to be in configuration 
$\sigma(\vec x,t),\varphi(\vec x,t)$.
The vacuum at $\sigma_-$ is metastable and will decay via tunneling, and the ground state energy is $E=V(\sigma_-)=0$.One then solves a stationary functional Schrodinger equation [23]
\[ H \psi = 0; \quad \mbox{where} \quad \pi_\sigma = -i \frac{\delta}{\delta\sigma}, \quad
\pi_\varphi = - i \frac{\delta}{\delta\varphi} . \]

Assume the WKB ansatz $\psi = \frac{e^{iS_0\hbar}}{\sqrt{S'_0}}  \psi[\varphi]$ 
where
$S_0$ is the action of the bounce $\sigma$ (following [7] Vilenkin) and the 
solution
for $\sigma$ in the thin wall approximation is: ( [1,7,8,10], etc) 
 $\sigma(r) = - a \, \tanh \mu(r - R)$. Here
$\mu = a \sqrt{\lambda}$, $r = \sqrt{\tau^2 + x^2}$. The thickness of the
wall is $L \approx \frac{1}{\mu}$. R is the radius of the bubble at the moment
of nucleation and $\mu R \gg 1$, where $R = \frac{3S_1}{\epsilon}$, and
$S_1 = \int^a_{-a}\sqrt{2V(\sigma)} d\sigma$ is the action for 1-dimensional instanton.
When solving for the $\sigma$ field, one reduces the problem to 
a $1-\dim$ 
$QM$ problem by integrating out the other degrees of freedom (x-dependence)
and ending up with:
$\int d^3 x \sqrt{-g} \widehat{H}(\sigma) = \widehat{H}(s), s$-the parameter that
plays the role of a time-coordinate.  In this case (see Vashapati and Vilenkin [7]):
\[ S_0 = -2 M_s U(s), \quad M_s = \int d^3 x \left( \frac{\partial\sigma}{\partial s}\right)^2, \quad
U(s) = \int d^3 x(\nabla\sigma)^2 + V. \]
From the $O(4)$ symmetry under consideration, we have 
\[ \sigma = \sigma(x^2 + \tau^2) = \sigma(x^2 - t^2), \begin{array}{ll}
\tau \rightarrow \mbox{euclidean time}& \\
t \rightarrow \mbox{real time} &
\end{array}\] 
The parameter ``$s$''is naturally chosen to be the variable : $s^2 = x^2 + \tau^2 = x^2 - t^2$; $t$ is defined as $\frac{ds}{dt} = \frac{1}{M_s}\frac{dS_0}{ds}$; $E = 0$; and
$\tau$ is the Euclidean time .  The tunneling rate for
$\sigma$ is $\gamma = e^{-S_{0_E/}/{\hbar}}$ where $S_{0_E}$ is the action in Euclidean time (and where $t \rightarrow - \tau$). The well known result for the bounce action [1,19], is: 
$S_{0_E} = 27 \pi^2 S^4_1/2\epsilon^3$. 
\item[{\bf 2.3 }] {\underline{\bf Coordinates:}}  Taking into consideration the 
$O(4)$ symmetries we will use the following coordinate system ([9])
\[ \left\{ 
\begin{array}{lll}
r_m = \rho\; \sinh \chi & \\
t_m = \rho \cosh \chi    &  \\
\Omega = \Omega                  & \\
0 < \rho < \infty, 0 < \chi < \infty   
\end{array} \right.\]
where $ r_m, t_m $ are the Minkowski radius
and time.  The Minkowski line
element written in the new coordinates is:
$ ds^2 = - d\rho^2 + \rho^2 [d\chi^2 + \sinh^2\chi d^2r] $.
In the Euclidean region, under the barrier, the coordinates are: 
\[ \left\{ \begin{array}{lll}
\rho_E \rightarrow -i \rho & \\
                           & ds^2_E = d\rho^2_E + \rho^2_E(d\chi^2_E + \sin^2\chi_E d^2r)\\
\infty>\rho_E>0, \chi_E \rightarrow i\chi 
\end{array} \right. \]

Then the future Lorentz timelike region in the above coordinates is the Milne universe and
$\sigma = \sigma(\rho)$.[24]. This variable $\rho$ is, clearly, the variable $s$ of the previous Section 2.2.
\item[{(\bf 2.4)}] {\underline{\bf Boundary Condition and Particle Creation
 in Functional}}\\
 {\underline{\bf Schrodinger and 2nd Quantization Picture.}}

The field $\sigma$ interpolates under the barrier between the 2 values,
$\sigma_-, \sigma_+,$ and $\infty >\rho_E > 0$, $\rho^2_E = x^2 + \tau^2$.
We want $\sigma$ to reach false vacuum when $\rho_E \rightarrow \infty$,
$\tau \rightarrow - \infty$, and the fluctuation field  $\varphi$ to die out there (i.e. at
$\rho_E \rightarrow \infty$ the field is in the false vacuum $(\sigma_-)$
ground state), i.e. $\phi \simeq e^{ikx+\omega_-\tau}$ where 
$- \infty < \tau < 0$,
or equivalently $\phi \sim e^{- \omega_-\rho_E}$ for $\rho_E \rightarrow \infty$.
The $In$ and $Out$ frequency is defined as usual by $\omega_{+/-} = [k^2 + m^2(\sigma_{+/-})]^{1/2}$.The field reaches the turning point at $\tau = t = 0$ and from then on has 
damped oscillations around the true vacuum $\sigma_+$ outside the barrier,
i.e.\\
\\
at  
$\quad\quad\quad\quad \rho_E = \rho = 0 (\tau = t = 0), \sigma \rightarrow \sigma_+ \;
(\mbox{where} \; 0 < t < \infty). $\\
\\
When $t \rightarrow \infty,$\\ 
\[ \left\{ \begin{array}{ll}  
\varphi \sim e^{ikx-i\omega_+ t}&\\
\sigma = \sigma_+ &
\end{array} \right. .\]
Therefore we analytically continue the solution for $\varphi$ at the turning point 
$\rho_E = \rho = 0$, and compare it with what should have
been a positive frequency wave in Minkowski region around $\sigma \simeq \sigma_+$,
i.e. with 
\begin{equation}
\varphi \sim e^{ikx-\omega_+t}
\end{equation}
In general the analytically continued solution differs from (2.2) in that it
contains a mixing of positive and negative frequency.  This is to be 
expected since the analytically continued solution from under the barrier
contains the dynamics of the $\sigma$-field interpolating between 2 values
and thus the phase change is reflected and carried over when one continues
that solution outside, in Minkowski region.  These excitations are interpreted as
particle creation.

In the second quantization picture, the existence of a mass-squared term that  
is ``time($\rho$)-dependent'' would produce the phenomena of particle production.
Thus $\varphi$ is in a'' time-dep''. background of $\sigma$ and it has been
shown by [7,9] that both, functional Shrodinger and
2nd quantization are equivalent methods in the calculation of particle
creation.
\end{itemize}
\section{\underline{Solution for the Fluctuation Field $\Phi(\rho, \chi, \Omega)$}}
\setcounter{equation}{0}
 In the Euclidean region,the equation for the fluctuation field\footnote{Note that $\Phi$ is now used to denote the fluctuation field
$\varphi$ of the previous sections.}, which is derived from the hamiltonian in Section 2.2 is
\begin{equation}
[\partial^2_{\rho_E} + \frac{3}{\rho_E} \partial_{\rho_E} + 
\frac{1}{\rho^2_E} \cdot \widehat{L}^2_{E} - m^2(\sigma)]\Phi = 0 
\end{equation}
where
\[ \widehat{L}^2_{E} = \frac{1}{\sin^2\chi_{_E}}
\partial_{\chi_{_E}}\left(\sin^2\chi_{_E}\partial_{\chi_{_E} }\right) 
+ \frac{\nabla^2_\Omega}{\sin^2\chi_{_E} }\]
In the Minkowski region the operator is $-\widehat{L}^2 = \frac{1}{\sinh^2\chi}
\partial_{\chi}\left(\sinh^2\chi\partial_{\chi}\right) 
+ \frac{\nabla^2_\Omega}{\sinh^2\chi}$. 
Also in the Euclidean region $p$ takes discrete values from 0 to $\infty$ while in the Minkowski region the spectrum becomes continuous [21]. 
We can write
\begin{equation}
\Phi(\rho_{_E}, \chi_{_E}, \Omega) = \frac{\psi(\rho_{_E})}{\rho^{3/2}_{_E}} \cdot
F_{p\ell m}(\chi_{_E}, \Omega)
\end{equation}
where the operator $\widehat{L}^2_{E}$ has eigenfunctions
$F_{p \ell m}$ such that
\[ \widehat{L}^2_{E_{\chi,\Omega}} (F_{p \ell m}) = 
\lambda'_p F_{p \ell m}(\chi_{_E},\Omega), \quad 
\lambda'_p = 1 + p^2 \]
Then, after replacing (3.2) in (3.1) we obtain:
\begin{equation}
\left[ \partial^2_{\rho_{_E}} + \frac{(p^2 + 1/4)}{\rho^2_{_E}} - 2 \mu^2
[3\tanh^2[\mu(\rho_{_E} - R)] - 1] \right]\psi = 0
\end{equation}
We have used,
\[ m^2[\sigma(\rho_{_E})] =  m^2(\rho_{_E}) = V''[\sigma(\rho_{_E})] = 
2\mu^2[3\tanh^2[\mu(\rho_{_E}-R)] - 1 ] \]
and 
\[ \sigma(\rho_{_E}) = -a\tanh[\mu(\rho_{_E}-R)] \]
as shown in Section (2.2)
Equation (3.3) is a Schrodinger equation of a particle of zero energy in a potential
$(E-V_0) = \frac{(p^2 + 1/4)}{\rho_{_E}} - m^2(\rho_{_E})$.Thus the inverted potential $V_0$ is
\begin{equation}
V_0 = - \frac{(p^2 + 1/4)}{\rho^2_{_E}} + m^2(\rho_{_E}) \quad (\mbox{for}\; 
E = 0 \; \mbox{particles}) .
\end{equation}

This potential is shown in Figure 3.1.  In the thin wall approximation we have $\mu R \gg 1$
where $\mu = \frac{1}{L}$, and $L$-thickness of the wall
\vspace*{2in}\\
Fig 3.1  The potential $V[\rho]$ for the fluctuation field of equation (3.3).

Looking at the 4 regions we can approximate the potential (fig. 3.1) by (later we show that the error made in this approximation is quite negligible)
\begin{eqnarray}
V_0 = \left\{ \begin{array}{llll}
4\mu^2 & R + L< \rho_{_E} < \infty & (III) \\
-\alpha_3 \left[ x - \frac{\alpha_2}{\alpha_3}\right]^2 - k 
& \left|x-\frac{\alpha_2}{\alpha_3}\right| < L & (II) \\
4\mu^2 & \rho_{_0} \leq \rho_{_E} \leq R - L & (I) \\
4\mu^2 - \lambda_{P/\rho^2_{_E}} & 0 \leq \rho_{_E} \leq 2\rho_{_0} & (H) 
\end{array}\right.
\end{eqnarray} 
where:
\[ \left\{ \begin{array}{lllllll}
x = (\rho_{_E} - R), & \alpha_2 = \frac{\lambda p}{R^3}, & 
\alpha_1 = (\frac{\lambda p}{R^2} + 2 \mu^2), & \alpha_3 = (\frac{3\lambda p}{R^4} -6 \mu^4)\\
k = \alpha_3 [ \frac{\alpha_1}{\alpha_3} - \frac{\alpha^2_2}{\alpha^2_3}], &
\omega_p = 2 \mu, & \lambda_p = p^2 + 1/4 
\end{array} \right. \]
 The regions of interests are region $II$, and $H$. We find the solutions in those regions and match them. However, we included regions $III$ and $I$ as separate, in order to indicate that the potential is nearly constant there. These latter regions were thus used to check the right asymptotic behaviour of the solutions found in the regions of interest.The turning points are:
\begin{equation}
x_{1,2} = \frac{\alpha_2}{\alpha_3} \pm \sqrt{\frac{-k}{\alpha_3}}, \quad
\rho_0 = \sqrt{\frac{\lambda_p}{4 \mu^2}} .
\end{equation}
 
We will find the exact solutions in the 4 regions
and match them and their first derivatives at the turning point.  Note that
if $\lambda_p \gg (n \mu R)^2$ for large values of n, then the $V_0$ defined above or the originial potential, to a very good accuracy, simply becomes 
$V_0 = - \frac{\lambda_p}{\rho^2_E} + 4 \mu^2$, i.e.
 ``$\frac{\lambda_p}{\rho^2_E}$''-term dominates the $\tanh^2 \mu\rho$ and thus the distinction between the first three regions dissapears.
In this case our result recovers that of the Kyoto group [to within a prefactor] [1,9] as it
should since they take the mass term, $m[\rho]$ in the equation 3.4 for the potential, to be a constant. For sufficiently small values of $p$, one has
$\lambda_p \leq (\mu R)^2$,in which case  the $\rho_E$ dependence of 
$m^2(\rho_E)$-term should not be ignored  since it produces interesting effects.
The solutions for the 4-regions are:

{\underline{\bf{Region III:}}} 
\[ R + L < \rho_{_E} < \infty, 
\quad\quad 
\psi_{III} = A_3 e^{- \omega \rho_{_E}}, 
\quad\quad
\omega = 2\mu \]
which dies out at $\rho_e \rightarrow \infty$ (approaching the false vacuum
from under the barrier) as it should from the boundary condition.
  
{\underline{\bf{Region II:}}}
\[ R - L \leq \rho_{_E} \leq R + L, \quad\quad x = \rho_{_E} -R\]
The equation to solve is:
\begin{equation}
[ \partial^2_x - \left\{ \alpha_3 [x - \frac{\alpha_2}{\alpha_3}]^2 
+ k \right\}]\psi_{II} = 0 
\end{equation}
The solution to (3.7) are parabolic cylinder functions $U(b,z),; U(b,-z)$
(or Debye function: $U(b,z) = D^{(z)}_{-b-\frac{1}{2}}$ i.e.
\begin{equation}
\psi_{II} = A_2 U(b,z) + B_2 U(b, -z) 
\end{equation}
\[ \mbox{where:}\quad
b = \frac{1}{4} \frac{k}{\sqrt{\alpha_3}}, \quad
z = \alpha^{1/4}_3 (x - \frac{\alpha_2}{\alpha_3}) \]
The reason we choose $U(b,\pm z)$ is because around the turning points
$\{x_1,x_2\}$ they reduce to Airy functions i.e. have the right behavior
and are solutions there too. (If we were to use $WKB$ we would have Airy
functions by approximating $V_0$ with a straight line around the turning 
points.)
The asymptotic form for $U(b,z)$ when $|b|$ is large (in our case that is
true, $|b| \simeq \mu^2 R^2, z_{1,2} \approx \frac{\sqrt{2}}{\mu R}$
becomes:
\[ U(b, \pm z) \sim 2^{- \frac{1}{4} - \frac{b}{2}} \Gamma \left( \frac{1}{4} - \frac{b}{2} \right)
\left( \frac{t}{\xi^2 - 1} \right)^{1/4} \left\{ \begin{array}{ll}
A_i & (t)\\
B_i & (t)
\end{array}\right. \]
where:
\[ \xi = \frac{z}{ 2\sqrt{|b|} }, \quad t = (4|b|)^{1/4}\tau, \quad \tau = arc \cosh \xi - \xi\sqrt{\xi^2-1} \]
Thus $\psi_{II}$ is regular at the turning points with the right behavior.
When $b \gg z^2$ (in our case $b \simeq 2 \mu^2 R^2, z^2 \simeq \frac{2}{(R\mu )^2})$,
we have: 
\begin{eqnarray}
U(b, \pm z) \simeq \frac{\sqrt{\pi}}{2^{\frac{b+1}{2}}\Gamma(\frac{3}{4} + \frac{b}{2})}
e^{\mp \ell z + v_{1,2}}
\end{eqnarray}
where  
\[ \ell = \sqrt{|b|}, v_{1,2} \simeq \frac{z^3}{2\ell} - \frac{z^4}{\ell^4} 
+ \ldots \rightarrow 0 . \]   
When $z^2 \gg 1$ (region III), $U(b, \pm z) \sim e^{- z^2/2}$ i.e. dies o ut.
Thus $U(b, \pm z)$ has the right behavior in agreement with our boundary
conditions.

{\underline{\bf Region(I):}}  $\rho_0 \leq \rho_{E}$, very flat.  \\
Solutions are:
\begin{eqnarray}
\psi_I = A_1 e^{- \omega \rho_{E}} + B_1 e^{\omega\rho_{E}}
\end{eqnarray}

{\underline{\bf Region(H):}} $0 \leq \rho_{E} \leq \rho_0$\\
Equation (1.3) in this region becomes:
\begin{eqnarray}
[ \partial^2_{\rho_{E}} + \frac{\lambda_p}{\rho^2_{E}} - 4 \mu^2]\psi_A = 0 
\end{eqnarray}
This can be solved exactly in terms of Bessel function.  The solution to (1.10)
is
\[ \psi_A = \rho^{1/2}_E {\cal C}_{\pm i_p} (2 \mu \rho_E), \quad
{\cal C}_{i_p}-\mbox{any modified Bessel function} . \]
We choose Kelvin and Neuman functions: $K_{i_p}(\rho_E),I_{-i_p}(\rho_E)$. When
$\rho_E \rightarrow 0$ and $\rho_E \rightarrow - i\rho$ they become Hankel
functions, i.e. regular at the origin $\rho_E = \rho = 0$ and with the right
behavior when $\rho \rightarrow \infty$.
\[ \rho^{1/2} H_{i_p} (2 \mu\rho) \longrightarrow_{\rho\rightarrow\infty}
\sqrt{\frac{\pi}{2\mu}} e^{[i\omega\rho - (ip + \frac{1}{2})\frac{\pi}{2}]} \]
in the asymptotically flat region, i.e. recover Minkowski vacuum at the
true vacuum side of the barrier at late times.  Thus:\\
\begin{eqnarray}
\psi_A = C_2 K_{i_p}(2 \mu \rho_E) + C_1 I_{i_p}(2 \mu \rho_E)
\end{eqnarray}
{\underline{\bf Matching at $x = x_1$}}, 
$\quad x_1 = \rho_{1E} + R+ \frac{\alpha_2}{\alpha_3} = \sqrt{\frac{-k}{\alpha_3}}\quad$
$\left\{ \begin{array}{ll}
\psi_{III}(x_1) = \psi_{II}(x_1) & \\
\psi'_{III}(x_1) = \psi'_{II}(x_1)
\end{array}\right.$ \\
Using the asymptotic form of parabolic cylinder functions $U(b, \pm z)$
(3.9) and after tedious algebra we get:
\begin{eqnarray}
\left\{ \begin{array}{ll}
2A_2 = e^{\theta_2 - \theta_1} A_3(1 + \frac{\omega}{q}) & \\
2B_2 = e^{\theta_2 - \theta_1} A_3(1 - \frac{\omega}{q}) &
\end{array}\right.
\end{eqnarray}
where 
$\theta_2 = \ell z_1 =\frac{1}{2}{\sqrt k} \Delta {\sqrt{\alpha_3}}$
where
\begin{eqnarray}
\left\{ \begin{array}{llll}
\Delta & = & \sqrt{\frac{-k}{\alpha_3}} \rightarrow \; \mbox{real}\;\;
z_1 = \alpha^{1/4}_3 \sqrt{\frac{-k}{\alpha_3}} = \Delta\alpha^{1/4}_3\\
\Delta & = & \sqrt{(\lambda_p + 2\mu^2 R^2 + \frac{\lambda^2_p}{\mu^4 R^4 - \lambda_p})
\frac{R^2}{(\mu^4 R^4 - \lambda_p)}}  \\
        & = & R \sqrt{ (\lambda_p + 2\mu^2 R^2 + \frac{\lambda^2_p}{\mu^4 R^4 - \lambda_p})
         \frac{1}{(\mu^4 R^4 - \lambda_p)}}  \\
\end{array}\right.
\end{eqnarray}
Define  $q=\frac{1}{2} {\sqrt{k}}$ 
Thus
\[ \theta_2 = iqx_1\]
when $\lambda_p \ll \mu^4 R^4$ (true in our case) then $\Delta \cong \frac{\sqrt{2}}{\mu} = \sqrt{2}L$
and $\theta_1 = 2 \mu (R + \frac{\alpha_2}{\alpha_3} + \Delta)$.

{\underline{\bf Matching at $x = x_2$}}, $x_2 = \rho_{2E} + R + \frac{\alpha_2}{\alpha_3} =- \Delta$.
In the same manner as above we have:
\[ \left\{ \begin{array}{ll}
\psi_I(x_2) = \psi_{II}(x_2) \\
\psi'_I(x_2) = \psi'_{II}(x_2)
\end{array}\right. \]
Then (proceeding in exactly the same way) and replace $\{A_2,B_2\}$ in
favor of $A_3$:
\begin{eqnarray} \left\{ \begin{array}{ll}
2A_1 = A_3 e^{-\theta_3 - \theta_1} \{ 2 \cosh (2 \theta_2) + \sinh (2 \theta_2)\cdot ( \frac{\omega^2 + q^2}{\omega q} ) \} \\
2B_1 = A_3 e^{+ \theta_3 - \theta_1} (\frac{\omega^2 - q^2}{\omega q}) \sinh (2\theta_2);  \theta_2-\mbox{imaginary}
\end{array}\right.
\end{eqnarray} 
where $\theta_3 = 2\mu(R + \frac{\alpha_2}{\alpha_3} + \Delta)$

{\underline{\bf Matching at $\rho_E =2 \rho_0$.}}
\[ \left\{ \begin{array}{ll}
\psi_A (\rho_0) = \psi_I(\rho_0)\\
\psi'_A(\rho_0) = \psi'_I(\rho_0)
\end{array}\right.    \]
\[ \psi_A = \rho^{1/2}_E \cdot [ C_2 K_{i_p}(2 \mu \rho_E) + C_1 I_{i_p}
(2\mu\rho_E)] \longrightarrow_{\rho_0 \gg 0} 
\sqrt{ \frac{\pi}{4\mu} }
[C_2 e^{\omega\rho_0} + C_1 e^{- \omega\rho_0}] \]
\begin{eqnarray}
\Longrightarrow \left\{ \begin{array}{ll}
A_I = C_2 \sqrt{ \frac{\pi}{4\mu} } & \\
B_I = C_1 \sqrt{ \frac{\pi}{4\mu} } &
\end{array}\right. 
\end{eqnarray}

The coefficient $A_3$ is found from the unitarity condition.  If
$R = |\frac{C_1}{A_3}|^2, \; T = |\frac{C_2}{A_3}|^2, \; R,T$ reflection
and transmission coefficient in Shrodinger picture in the Euclidean region
then $R + T = 1$ means $|A_3|^2 = |C_1|^2 + |C_2|^2$.  

Now that we have the solution we can analytically continue it through
$\rho_E = \rho = 0$ (turning point) in order to obtain the Bogolubov
coefficient.  This is done by replacing $\rho_E \rightarrow - i\rho$
and using the properties of Bessel functions.  We have:
$\;\;\;\;\;\;\;\;\; \psi^{(\rho_E)}_A = [C_2 K_{i_p}(2\mu\rho_E) + C_1 I_{i_p}(2\mu\rho_E)]
\rho^{1/2}_E .\;\;\;\;\;\;$\\
In the outside region, near $\rho \simeq 0$, that function is 
\[ \psi_A(\rho) = [a_1 H^{(1)}_{i_p}(2 \mu \rho) + a_2 H^{(2)}_{i_p}(2\mu\rho)]\rho^{1/2}. \]
But when we match $\psi_A(\rho_E \simeq 0)(*)$, we use:
$ \rho_E \rightarrow - i\rho)$.  Then;
\[ \left\{ \begin{array}{ll}  
K_{i_p} (- i 2\mu\rho) = \frac{\pi i}{2} e^{- \pi p/2} H^{(1)}_{ip}(2 \mu \rho) &\\
I_{i_p}( - i 2 \mu\rho) = \frac{1}{2} e^{\pi p/2}[H^{(1)}_{i_p}(2 \mu \rho) +
H^{(2)}_{i_p}(2 \mu \rho)] 
\end{array} \right. \]
and from (*) we obtain:
\begin{eqnarray}
\left\{ \begin{array}{ll}
a_2 = \frac{1}{2} C_2 e^{\pi p/2} & \\
a_1 = \frac{1}{2} e^{\pi p/2} [C_2 + \pi i \cdot e^{- \pi p} \cdot C_1]
\end{array} \right.
\end{eqnarray}

When $\rho \rightarrow \infty$,$\;$ $H^{(1)(2)}_{i_p}(2 \mu \rho) \sim
\sqrt{\frac{\pi}{4\mu\rho}} \exp [\pm 2 \mu i \rho \mp i(ip + 1/2)\pi/2]$.
Thus the 
l out $>$-state at $\rho \rightarrow \infty$
in the true vacuum, $\psi_A (\rho \rightarrow \infty)$ is a squeezed state,
a mixture of positive and negative frequencies:
\begin{eqnarray}
\psi_A (\rho \rightarrow \infty) \longrightarrow a_2 e^{-i[2 \mu\rho - \frac{i p \pi}{2} - \frac{\pi}{4}]}
+ a_1 e^{ i [ 2 \mu \rho - \frac{ip\pi}{2} - \frac{\pi}{4} ] }
\end{eqnarray}
>From relation (3.17), (3.18) (for a more rigorous treatment in
 2nd quantization,
Heisenberg picture, see Kyoto, Vilenkin [7,9]) we obtain the particle 
production rate:
\begin{eqnarray}
\;\;\;\;n_p &\! =\! & \frac{|a_1|^2}{\pi^2|a_2|^2}\\
& \! =\! & 
\frac{4 \omega^2 q ^2 \cos^2 (\Delta^2 {\sqrt{-\alpha_3}}) - 
\pi^2 e^{- 2\pi p - 2 \theta_3} \cdot \sin^2(\Delta^2 {\sqrt{-\alpha_3}})\cdot [\omega^4 + q^4_]}
{4(\omega^2+q^2)^2 \cos^2(\Delta^2 {\sqrt{-\alpha_3}}) + \sin^2(\Delta^2 {\sqrt{-\alpha_3}})\omega^2q^2}\nonumber
\end{eqnarray}  
$\underline{\quad\quad\quad\quad\quad\quad}$\\
{\small (*) The error made in approximating the real potential to $V_0$, around the
turning points is of the order $(\frac{1}{\mu R})^8$ and higher, i.e. quite
negligible.}
{\underline{\bf Notation:}  $\omega = 2 \mu, \; q = \frac{1}{2}{\sqrt{k}}, \; 
\theta_2 = \frac{ \Delta^2}{2} {\sqrt{\alpha_3}}$.  Both ; $\{ q, \Delta\}$ are real (for $\lambda_p \leq (\mu R)^4$).
So $\sinh(2 \theta_2) \rightarrow - i \sin (\frac{\Delta^2 {\sqrt{-\alpha_3}}}{2})$.
\[ \left\{ \begin{array}{llr}
q = \frac{1}{2R} \sqrt{ (\lambda_\rho + 2 \mu^2 R^2 + \frac{\lambda_p^2}{\mu^4 R^4 - \lambda_p )} }& \\
                                        & $\mbox{(see 3.14)}$ \\
\Delta = R\sqrt{ (\lambda_p + 2\mu^2R^2 + \frac{\lambda^2_p}{\mu^4 R^4 - \lambda_p})
\frac{1}{(\mu^4 R^4 - \lambda_p)}} & 
\end{array} \right. \]
     
To get to formula (3.19) we used (3.13)-(3.17).  The calculation and analytic
continuation are straightforward.  

Relation (3.19) contains all the features found before from other authors
(see [8,9]) i.e. the suppression for low and high energies, namely
\[ n_p \sim e^{- 4 \mu R} \;\; \mbox{for} \;\; p-\mbox{small},\;\; 
n_p \sim e^{-2 \pi p}, \;\; p-\mbox{large} \]
but it differs greatly in terms that we considered a dynamic background (time-depend. mass)
while others have considered a constant mass-term or a step-function, and
this dynamics of $m^2(\rho)$ revealed an interesting phenomena, resonant
particle production.

As noted before the case $\lambda_p > 4 \mu^2 R^2$ where $\mu R \gg 1$
corresponds to an almost constant mass term studied previously by Kyoto
group [8] and we have studied the case 
$0 \leq \lambda_p \leq \mu^2 R^2 \cdot n^2$,
$(n = 0,1,2,3, \ldots)$ when the time-dependence of mass is important
and cannot be neglected anywhere, i.e. when ``the curvature of the bubble''
becomes important.  In general, from expression (3.19) we find that particle
production is strongly suppressed by the factor 
$\exp[-2\theta_2] = \exp[-4\mu[R + \frac{\alpha_2}{\alpha_3} + \Delta]] $
since $\mu R \gg 1$ and, 
$\left\{ \frac{\alpha_2}{\alpha_3}, \Delta\right\} \sim \frac{1}{\mu}$
small.  Also for small momentum, i.e. $p$ small we recover the result of
the Kyoto group in the thin wall approximate case: $n_p \sim e^{- 4\mu R}$
but we have a different prefactor from them, we don't have $(\mu R)^4$-facts
they find, and our result is finite integrated over all momenta-$p$.

The most interesting feature that comes from taking into account the
time-dependence of the mass-term, i.e. the curvature of the bubble
(background) is that for momentum ``$p$'' comparable to the curvature scale
 of the
bubble - 
``$\mu R$'' we get a strong enhancement of particle production,
a resonance effect.  In this case, $\lambda_p \sim n^2\mu^2 R^2$, the 
exponential factor becomes approximately zero and $n_p \simeq \frac{(4\omega^2q^2\cos^2/(\Delta^2 {\sqrt{-\alpha_3}})}
{4(\omega^2+q^2)^2\cos^2(\Delta^2 {\sqrt{-\alpha_3}})+\omega^2q^2\sin^2(\Delta^2 {\sqrt{-\alpha_3}})}$.  In this case the very high momentum ($p \leq (\mu R)^2$, see figure 3.3) all the sinusoidal functions become hyperbolic (the argument becomes imaginary) and the particle production rate dies out very fast as expected.
\begin{itemize}
\item[{1)}] Thus, for momentum less than the scale of the bubble ``$n_p$''
is very small due to the factor ``$e^{-4 \mu R}$''.
\item[{2)}] For large momentum $(p \gg n \mu R)$ one gets again a very small
``$n_p$'' suppressed by the factor ``$e^{- 2 p \pi}$''.  In this case there
is only one region:  $V_0 = \frac{\lambda_p}{\rho^2_E} - 4 \mu^2$ and
$n_p \sim e^{- 2 p \pi}$, as previously found [8].
\item[{3)}] Only for momentum comparable to the scale of the bubble
$(p \approx n \mu R)$, ($n$ takes only a few integer values) one gets a
resonant particle production.  Formula (3.19) contains cases (1) and (2)
as limiting cases and those have been previously studied by Kyoto [8,9].
Our results agree with theirs within a prefactor.

The result is that particle production is suppressed in the thin wall case
 with the exception of those modes that can resonate
with the bubble, and those modes are not suppressed but enhanced (not shown
previously), i.e. modes with phase 
``$p \pi$'' $\approx n \cdot \theta_3 \cong n \cdot 2 \mu R$ can see an
almost transparent barrier.  If the barrier was time-dependent i.e.
$\mu R = \mu R(t)$ one would expect intuitively a larger range of modes to resonate
with the barrier.  This has to be investigated yet, (using for e.g. [10]
model) (``$n$'' is an integer.  Only a few values of ``$n_p$'' give a strong
resonance, because of the suppression term $e^{-2 \pi p - 2\theta_2}$
competing with the resonance effect when $p$ is large.  See (figure 3.2, 
figure 3.3).
\end{itemize}
{\underline{\bf Figure 3.1}}
Resonant Particle Creation Number for small values of the momentum.

\vspace*{1in}
{\underline{\bf Figure 3.2}}
Particle creation number for high values of the momentum.

\vspace*{1in}
\section{\underline{Discussion.}}
If we did not distinguish between fluctuation and classical fields, fluctuations
are the ones that contribute to the phase shift of the wave functional and
give rise to particle production at the same time.  Particle production
is calculated in real time, after
the analytic continuation of $DEP$ outside.  Thus it is the total phase
shift coming from tunneling/$DEP$ that gives rise to particle production.
Further, on
support of the above, notice that the parameter ``$s$'' plays the role of time.  
Under the barrier $\dot{s} < 0$, outside $\dot{s}^2 > 0$ where
\[ \int d^3 x H[\sigma(x,s)] = H(s) = \frac{1}{2M_s} \dot{s}^2 + U(s), 
\left\{ \begin{array}{ll}
s^2 = x^2 + \tau^2 \\
M_s > 0 
\end{array} \right. .\]
Also, in a $QM$ picture ``$s$'' is an operator, the coordinate of the particle
in the potential $U(s)$ thus the ``kinetic energy''-$(\dot{s}^2)$ is
negative under the barrier.  It has been shown that letting $s$ run from
$s \rightarrow \infty$, to $s \rightarrow 0$ under the barrier is 
equivalent to summing up over all instantons in the dilute gas approximation.
But instantons are paths in configuration space.  Thus ``time''-evolution
of the field (i.e. let ``$s$'' run from $\infty$ to 0) is equivalent 
to the trajectory over the ensemble of paths in configuration space.
(think of it as phase space).  The changing of sign of $\dot{s}^2$ from
$\dot{s} < 0$ to $\dot{s}^2 > 0$ would correspond to an inflection point
in this trajectory.  That is also the turning point from under the barrier
to outside.  Taking this statistical point of view one would go from Euclidean
to Lorentz region at the inflection point.  In general $s = s(t)$ is a 
complicated function of time, thus $t_E \rightarrow it$ does not always
give the analytical continuation from $DEP$ to real time, i.e. is not crucial
for tunneling, but if one can reduce the problem to a few degrees of
freedom ($s$-in our case) going from ``negative kinetic energy of these
degrees'' to ``positive kinetic'' i.e. from $\dot{s}^2 < 0$ to
$\dot{s} > 0$ is inevitable in the case of tunneling, i.e. its intrinsic
to this process.  The reason is because this parameter is defined in terms
of the hamiltonian and the field is not an eigenstate of the hamiltonian
in the case of tunneling .  Outside the barrier $\dot{s}^2 > 0$ and the
``time''-evolution for $0 < s < \infty$ gives the in-out states.  In this
approach one considers instantaneous tunneling in real time, i.e. at
$s = 0$ (in real time parameter), the field hops from the false vacuum to the other side of the
barrier (true vacuum) and at $s \rightarrow \infty$ settles in the true vacuum
(l out$>$-state).  The solution at the turning point $s = 0$ which is the
endpoint of $DEP$ is analytically continued thus the phase shift from
$DEP$.  So, particle production comes from the phase shift and the 
l in-out$>$ states
are obtained from the analytic continuation of the $DEP$ in Lorentz
region, i.e. in terms of ``$s''$-parameter.  The in-out states are the
evolution of $DEP$ when $s \rightarrow \infty, 0$.

Then one could study real tunneling through particle production methods. 
Knowing particle production
and the l out$>$-state one has all the information about the phase shift/i.e.
tunneling, and evoluting the lout$>$-state back in ``time''-$s$ one can
obtain the $DEP$ by analytically continuing at the inflection point
(turning point where $\dot{s}^2$ goes through zero).
 
Particle production calculated in our case, when the mass is time-dependent,
i.e. dynamic background of the bubble, exhibits a strong resonance effect
with the bubble for those modes that are comparable to the curvature scale
of the bubble (i.e. for 
$p \pi \simeq n \theta_3 \; \mbox{or} \; p \simeq n \mu R, n = 1,2,3 \ldots 10 \ldots)$.
When the modes have a high momentum they are exponentially suppressed by
a factor $Exp[-2\pi p]$.  This feature of exponential suppression by
$Exp[-2\pi p]$ or $Exp[-4\mu R]$ has been previsouly found by [8,9] when
they considered $m^2$-term to be a step function or a constant in each
region respectively.  In these last 2 limiting cases $m_p$ has a thermal
behavior as previously discussed by [8,9].  See figure 3.2, 3.3.

One interesting case, currently under investigation, is when the barrier
itself is time-dependent.  From naive arguments we would expect a larger
range of modes to resonate with the bubble (as argued by B\"{u}tiker and
Landuer).  However we may find a nontrivial answer, different from the
above, once that calculation is done.

The modes of the $\varphi$-field, if one considers those as the environment,
will have an effect on the $\sigma$-tunneling field (considered as the
system).  That can be found through calculating the noise and dissipation
kernels in the case when the $O(4)$ symmetry is broken i.e. when we have
more than one degree of freedom for $\sigma$, thus bringing correlation
of the $WKB$ branches under the barrier.  We hope to reveal these results
within the next year.

\vspace*{1in}
\noindent
Acknowledgements:

I owe a special debt of gratitude to two people who with their insights, guidance, continuous support and effort, have contributed to this work as much as I have. They are; Prof.B.L.Hu, my advisor in UMD where the idea of this project originated, and Prof.L.Parker, my current advisor in UWM, where this research was carried out and finalized, under his close supervision and collaboration.

I would also like to thank Dr.A.Raval for the interest he showed in my work, for checking the calculations as well as his participation with the editing. Many thanks to Joyce Miezin for her help with typing.
\pagebreak\\
{\underline{\bf References.}}
\begin{itemize}
\item[{1.}] S. Coleman, Phys. Rev. D.16, 1662 (1977), C.G.Callan and S.Coleman, Phys Rev D16,1762(1977).
\item[{2.}] J. L. Gervais and B. Sakita, Phys. Rev. D.16, 3507 (1977).
\item[{3.}] H. J. Vega, J. L. Gervais, B. Sakita, PRD 19, 604 (1979).
\item[{4.}] V. A. Rubakov, Nucl. Phys B245, 481 (1984).
\item[{5.}] T. Banks, C. Bender, T.T. Wu, PRD D8, 3366, (1973).
\item[{6.}] T. Vachaspati, A. Vilenkin, PRD D37, 898 (1988).
\item[{7.}] T. Vachaspati, A. Vilenkin, PRD D43, 3846 (1991).
\item[{8.}] Sasaki, Tanaka, PRD 49, 1039 (1994).
\item[{9.}] Hamazaki, Sasaki, Tanaka, Yamamoto, PRD 53, 2045 (1995).
\item[{10.}] E. Keski-Vakkuri, Per Krauss, PRD 54, 7607 (1996).
\item[{11.}] L. Parker, PRD 183, 1057 (1969).
\item[{12.}] L. Parker, Nature, 261, 20 (1976).
\item[{13.}] B. K. Berger, PRD 12, 368 (1975).
\item[{14.}] W. G. Unruh, PRD 14, 870 (1976).
\item[{15.}] B. L. Hu, PRD 9, 3263 (1974).
\item[{16.}] L. Parker, in Asymptotic Structure of Space-Time eds. F. P. Esposito
and L. Witten (Plenum, New York 1977).
\item[{17.}] B. L. Hu and D. J. O'Connor, PRD 36, 1701 (1987).
\item[{18.}] B. L. Hu, Class. Quan. Grav. 10, 593 (1993).
\item[{19.}] S. Coleman, `Aspects of Symmetry'.
\item[{20.}] Abramovitz, M. and Stegum. A., eds. (1985), `Handbook of 
Mathematical Functions' (New York: Dover).
\item[{21.}] N. D. Birrel and P.C.W. Davies, eds. (1982), ``Quantum Fields in Curved Space'' (Cambridge Monographs on Mathematical Physics).
\item[{22}] more commonly known as the bounce solution, i.e. the solution for $\sigma$ under the barrier that interpolates between the two values of $\sigma$ for the false and true vacuum
\item[{23}] Since there is one integration constant (the wall of the bubble) the wave-functional
$\psi$ is peaked around a 1-parameter family of solutions $\sigma(x,s)$
where the  $s$-parameter is taken to be the integration constant in the solution for $\sigma$.
\item[{24}] There is no Cauchy surface in Milne Universe to cover the whole of Minkowski space. For subtleties related to that (e.g.
discreet modes) see [9].  We will not mention those in this paper as they do not affect our result.

\end{itemize}

\resizebox{\hsize}{8.0in}{\includegraphics{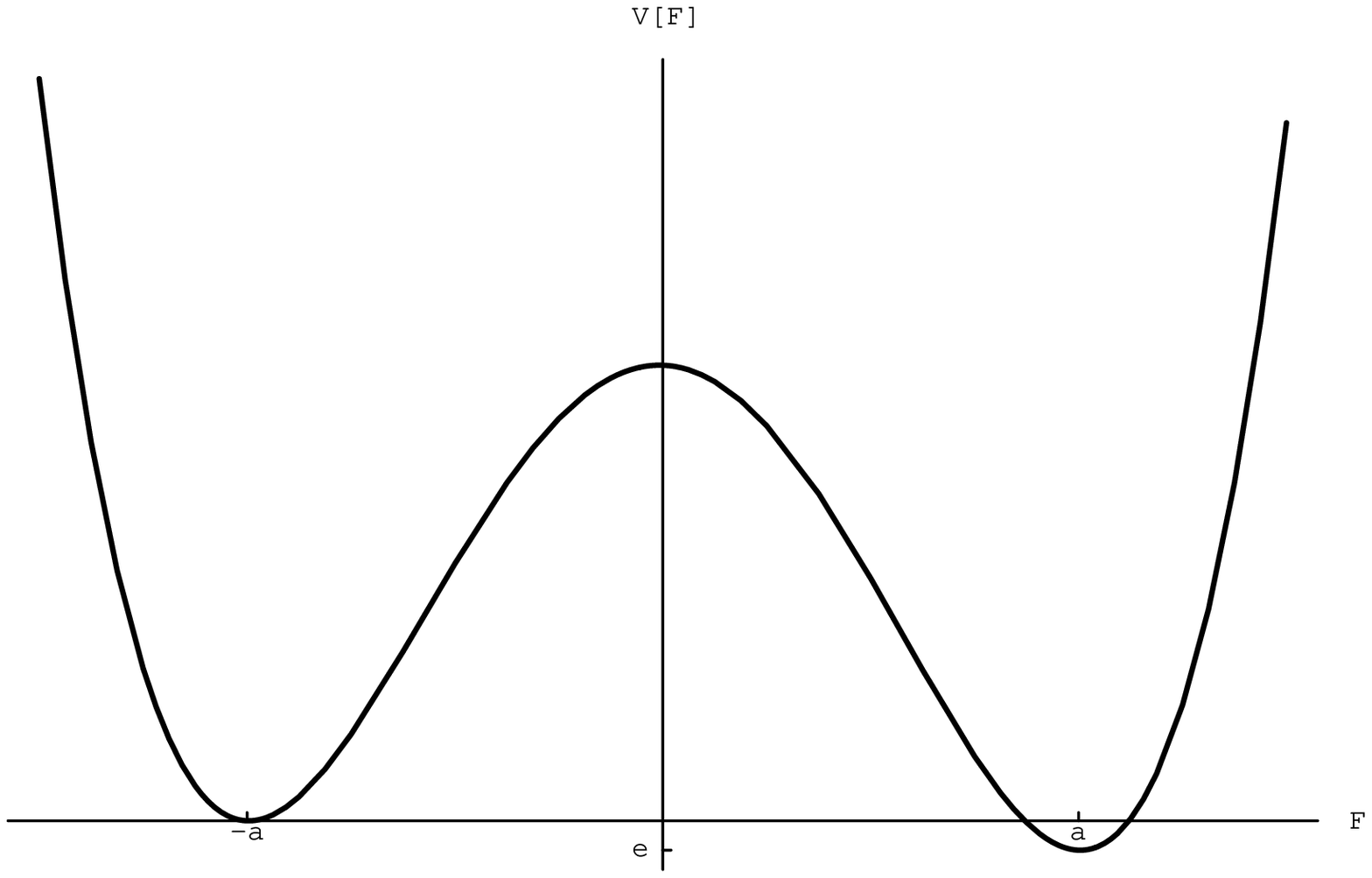}}
\includegraphics{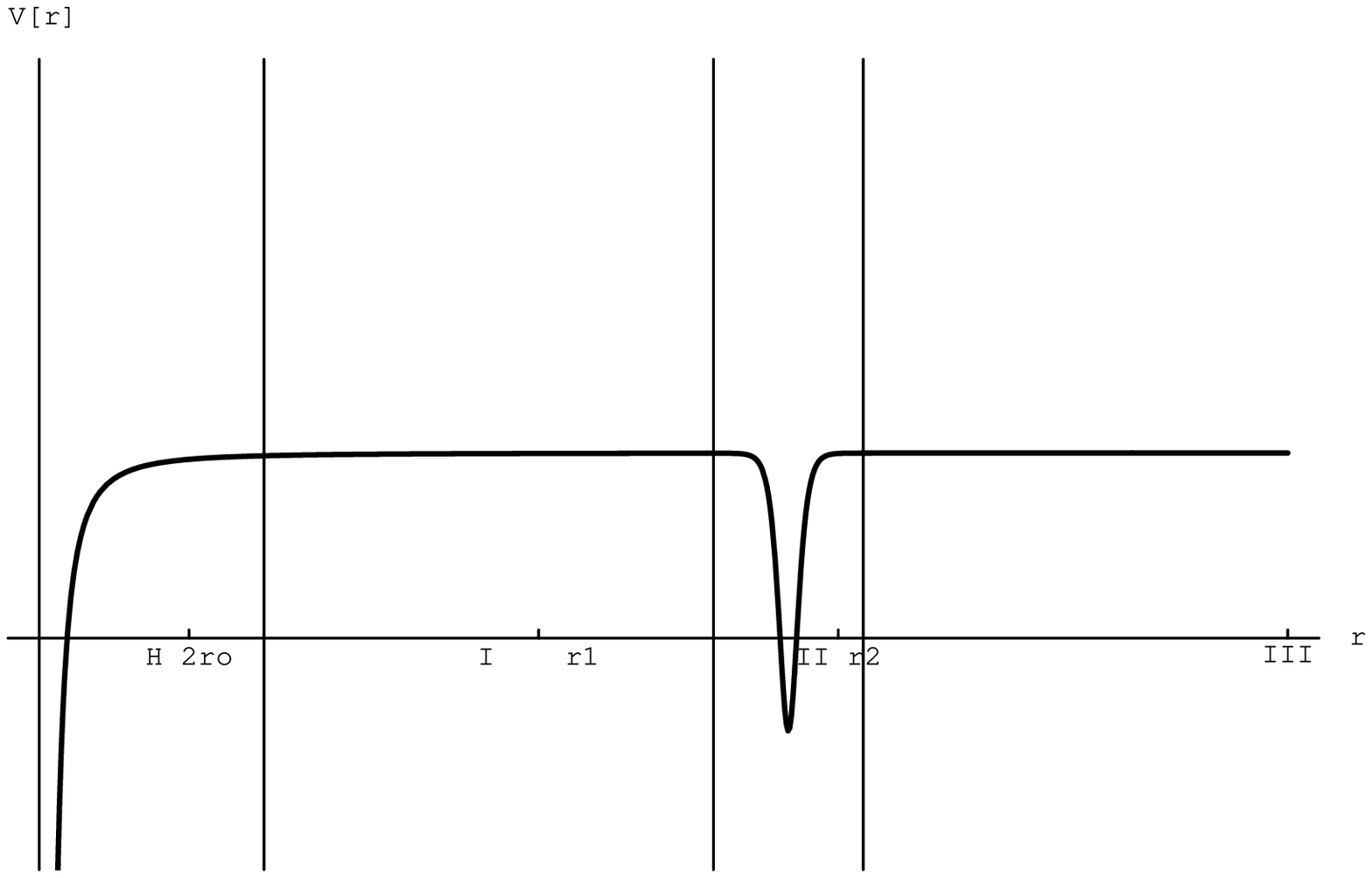}
\includegraphics{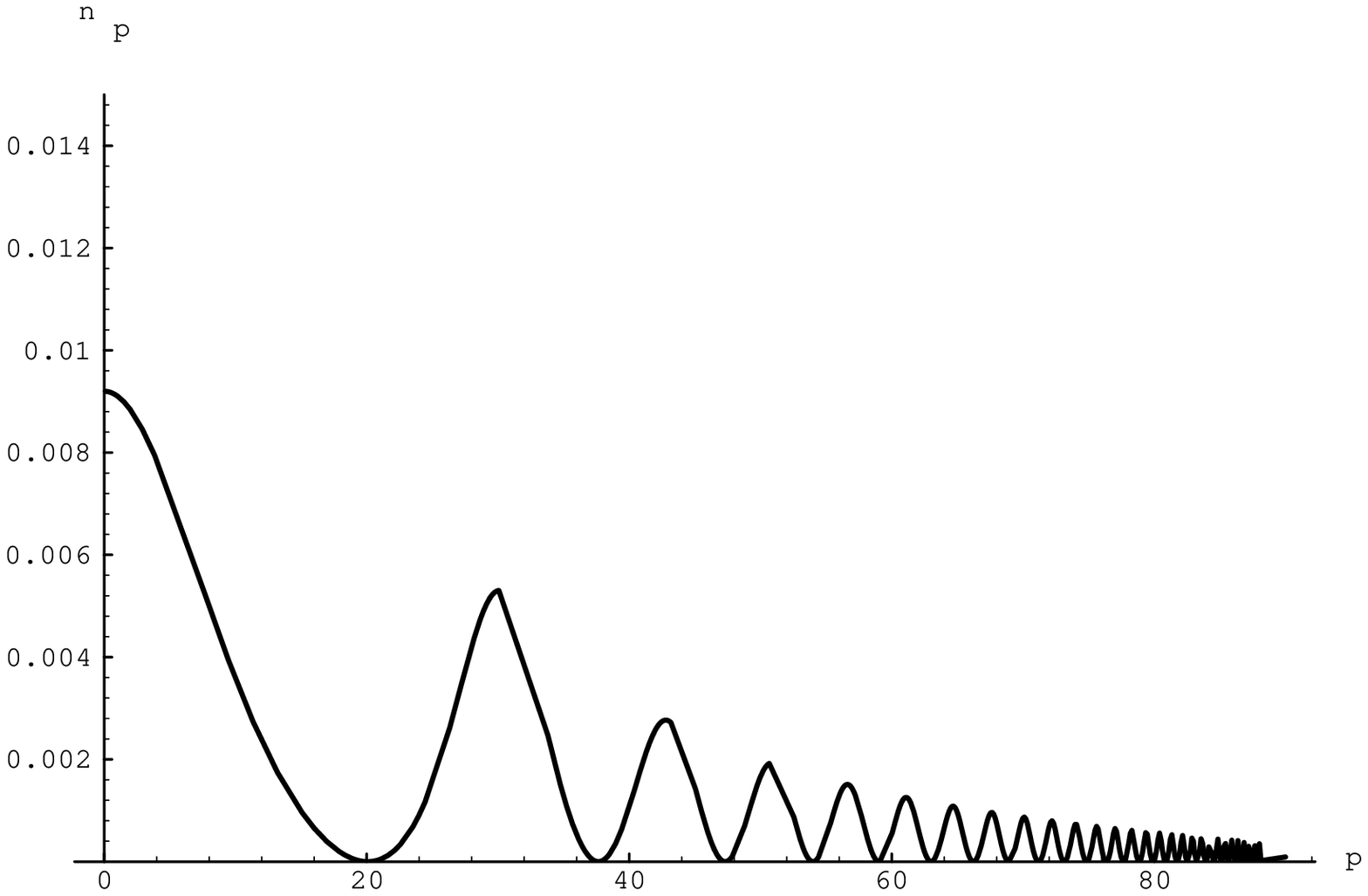}
\includegraphics{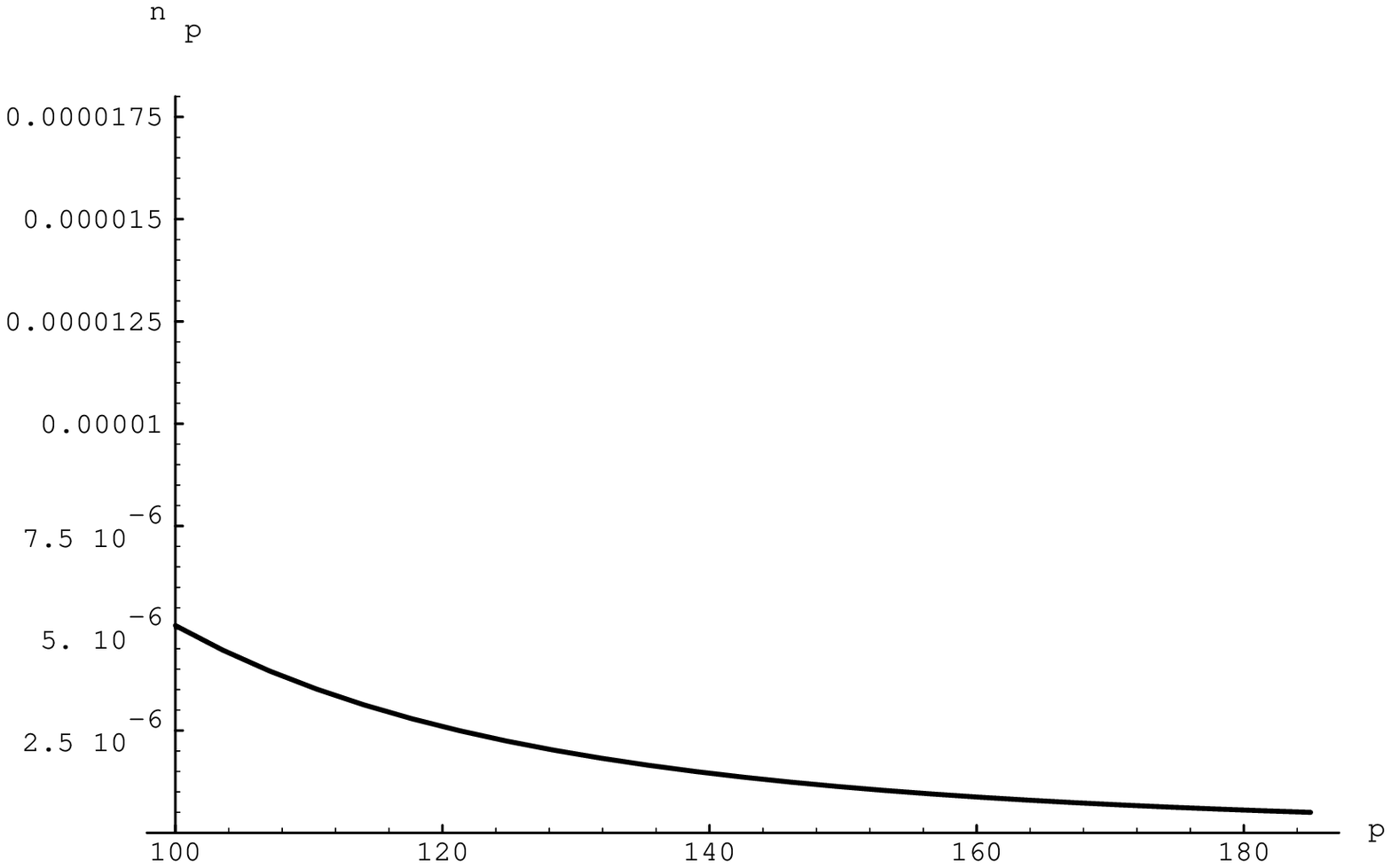}

\end{document}